\documentstyle[12pt]{article}

\newcommand{\be}{\begin{equation}}
\newcommand{\ee}{\end{equation}}
\newcommand{\beq}{\begin{eqnarray}}
\newcommand{\eeq}{\end{eqnarray}}

\catcode`@=12
\begin{document}
\begin{titlepage}

\rightline{\vbox{\halign{&#\hfil\cr
&NTUTH-96-03\cr
&March 1996\cr}}}
\vspace{0.2in}

\vfill

\begin{center}
{\Large \bf  Prospect for Heavy Quarks Lighter than {\boldmath $M_W$}  }

{\Large \bf                         at Tevatron and LEP II
\footnote
{
Talk presented at the JINR-ROC Symposium, 
June 26 -- 28, 1995, Dubna, Russia.
}
}
\vfill
        {\bf George Wei-Shu HOU}\\
        {Department of Physics, National Taiwan University,}\\
        {Taipei, Taiwan 10764, R.O.C.}\\
\end{center}
\vfill
\begin{abstract}
As the excitement surrounding the heavy top quark discovery
subsides, while the expectation for LEP II physics gathers,
it is a good time to sit back and reflect
on whether energy regions available to us have been fully explored.
We emphasize that a loophole exists where heavy quarks, perhaps the actual
top quark itself, could still be hidden below $M_W$.
This would typically involve scalar induced decays
of the heavy quark, and could be realized in models with more than
one Higgs doublet, e.g. MSSM.
We illustrate such mechanisms with
two Higgs doublet models, the addition of singlet quarks,
as well as reconsidering a fourth family of quarks and leptons.
Curiously, the present $R_b$--$R_c$ problem may be a harbinger
of such scenarios.
Given that LEP-II would be running soon,
and in view of the large amount of data that the Tevatron has collected,
we urge our experimental colleagues to conduct a critical analysis.

\end{abstract}
\vfill
\end{titlepage}

\section{Present Data and the Flavor Problem}

All experimental data seem to be fully consistent with the
Standard Model with 3 fermion generations (SM3).
It is truly remarkable that SM3 could
account for a very wide range of data.
Not only LEP precision tests fail to  reveal new
effects (except $R_b$--$R_c$, see below),
the flavor related parameters such as
$m_t \cong 180$ GeV
and $V_{cb} \simeq 0.8 \lambda^2$,
$V_{ub} \simeq 1.3 \lambda^4 e^{-i\delta}$
($\lambda \equiv V_{us}$
and $\delta$ is the CP violating phase in SM3)
could numerically account for subtle effects such as
$\Delta m_K$, $\Delta m_B$,
$\varepsilon_K$, $\varepsilon^\prime/\varepsilon$
and $b\to s\gamma$.

Remarkable as it is, we should recognize that,
``accounting" is one thing, but a deeper ``understanding"
({\it e.g.} gauge invariance as a dynamical principle)
of all this is still lacking.
To gain some perspective on the parameters of SM3, note that
\vskip -0.01cm
\begin{tabular}{crlccl}
gauge: & \ & $g_1$, $g_2$, $g_3$ &&& ($\Longrightarrow$ unification?) \\
symm. breaking:  & \ & $v$,
                        \fbox{\fbox{$m_H$}} &&& $\Longrightarrow$ LHC! \\
Flavor: & \ & $m_e,\ m_\mu,\ m_\tau$ &&& \\
              & \ & $m_u,\ m_c$, \fbox{$m_t$} &&& $\Longleftarrow$ 1994-95 Tevatron \\
              & \ & $m_d$, $m_s$, $m_b$     &&& \\
              & \ & $s_1$, $s_2$, $s_3$, \fbox{\fbox{$\delta$}} &&& $\Longrightarrow$ B Factory \\
                                                                                    \hline
\end{tabular}
\vskip 0.2cm
\noindent
In all,  \underline{13} out of 18 parameters are in Flavor Sector.
Furthermore, this number could easily multiply if more fermions are discovered.
If gauge couplings $g_i$ are basically understood,
while symmetry breaking is partially understood,
in comparison, FLAVOR is NOT UNDERSTOOD!
This constitutes the Flavor Problem.
Many questions are contained here:
\begin{itemize}
\item Why 3?
\item Why mass and mixing hierarchies?
\item Why $m_t \gg m_b$?
\item For weak bosons and the top quark, $M_W$, $M_Z$, $M_H$, $m_t \sim v$.
This appears to be natural since it is ``normal" to expect particles
to have mass of order the dynamical scale (e.g. hadrons in QCD).
The mystery is then, why all fermions (except the top) behave as
$m_f \ll v$? They appear more like zero modes at the weak scale
(see Fig. 1).
\end{itemize}
Together with our limited understanding of symmetry breaking,
the last point suggest that one could well have 
more states around the weak scale $v$.
These could be the 4th generation (SM4) fermions,
extra Higgs bosons, or the appearance of exotic (nonsequential) fermions.
What do we know about them? Perhaps the discovery of new particles
at the $v$ scale may provide us with better
understanding of the Flavor Problem.

It is useful to remind ourselves the hard facts at hand:
\begin{itemize}
\item A heavy quark weighing 180 GeV has been found during 1994-95
        \cite{Wang}.
However, {\it it is not yet proven to be the top},
since most heavy quark decay scenarios contain
$b$ final states. It could well be
the 4th generation $b^\prime$ or $t^\prime$ quark, or some exotic quark.
\item If a 4th generation exists (call it $E$, $N$),
then $m_E$, $m_N > M_Z/2$.
\item By same token, there is a firm lower bound of $M_Z/2$ for
$m_t$, $m_{b^\prime}$ and $m_{t^\prime}$.
\end{itemize}
In contrast, the limit of $m_t > 62$ GeV from ``$\Gamma_W$" measurement
by CDF is not firm since it assumes $V_{tb} = 1$.
The 1989-91 limit of $m_t > 91$ GeV by CDF
and 1994 limit of $m_t > 130$ GeV by D0 are not firm because
the SM3 value of BR$(t \to \ell \nu + X) \simeq 1/9$ is assumed.

Our theme, therefore, is that heavy quarks ($t$, $b^\prime$, $t^\prime$
or exotics) COULD lurk/hide below $M_W$
IF hadronic decays predominate the decay rate \cite{Hou1994}.
The existence of a heavy quark $Q^\prime$ with $m_{Q^\prime} < M_W$,
together with the observed heavy quark $Q$ with $m_{Q} \simeq 175$ GeV,
imply that there would be a lot of fun ahead of us
at both the Tevatron and LEP II.

\section{Mechanism to hide Top below $M_W$}

As an illustration of how the Tevatron experiments could have missed
heavy quarks with $m_Q < M_W$, let us discuss the
case for a ``light" top quark \cite{Hou1994}.
We define the top quark as the doublet partner of the $b$ quark.
For $m_t < M_W$, the SM3 decay chain is $t\to bW^* \to bf\bar f^{\prime}$,
where $W^*$ is virtual (see Fig. 2).
This process is suppressed by a propagator and coupling factor
$g^4/(q^2 - M_W^2)^2$ (i.e. still ``remembering" 4-Fermi interaction)
as well as 3-body phase space.
If some boson $X$ exist and couples to $t$-$q$ quarks with strength $\lambda_{tqX}$,
then, if $m_X < m_t$, this induces a 2-body decay.
If $\lambda_{tqX}$ is not too small compared to $g$ (the weak gauge coupling),
and if $X\to$ hadrons (on-shell $X$ decay),
then top decay could be dominated by hadronic final states due to $X$ production,
and it would be difficult for CDF/D0 to tell from multijet background.
Note that this cannot occur in SM3, since $tcH$ coupling does not exist.
However, $X$ could be exotic scalars such as $H^+$ or $h^0$.
These could arise from minimal SUSY (MSSM), where the scalars could
also be the top squark $\tilde t$ while $q$ above becomes
the chargino $\chi^+$ or neutralino $\chi^0$.
The simplest construction, however, would be the addition of
an extra scalar doublet.
In the following, we shall mainly use
two Higgs doublet models (2HDM) as a means of illustration,
before turning to more elaborate models.
For sake of space, we touch on salient features without giving any detail.

\section{Two Higgs Doublet Models}

The ``standard" 2HDMs invoke the natural flavor conservation condition (NFC)
of Glashow and Weinberg, where each type of fermion charge
receive mass from one Higgs doublet only.
The mass and Yukawa matrices are then simultaneously diagonized,
and FCNC couplings of neutral scalars are absent by construction.
For Model I, both $u$ and $d$ quarks receive mass from the same doublet,
while for Model II, they receive mass from different doublets.
The latter is popular because it naturally arises in MSSM,
and has been rather well studied.

It is possible to foresake the NFC condition by assuming
some approximate (global) flavor symmetry in the Yukawa couplings
to protect from low energy FCNC constraints.
This was noted already some time ago \cite{CS},
and has been discussed widely recently.

Let us see how a light top could be hidden in these models.

\subsection{Model I: $t\to bH^+$}

In this model, the $H^+$ coupling is
\begin{equation}
  \frac{\sqrt{2}}{v}\, V_{tb}\,
     \bar t \left(\cot\beta\,m_t L + \cot\beta\,m_b R\right) b + h.c.,
\end{equation}
where the parameter $\cot\beta = v_2/v_1$
is mainly constrained by $\Delta m_K$ and $\Delta m_B$.
The coupling could evade the stringent $b\to s\gamma$ constraint
by a cancellation mechanism between 
$H^+$ and $W^+$ effects \cite{HWil}.
It evades direct CDF search for $t\to bH^+$ as follows.
CDF finds \cite{tbH} that if $H^+\to \tau^+\nu$ is close to 100\%,
the entire region of $m_{H^+} < m_t < M_W$ is ruled out.
However, if it falls below 50\%, then the entire region becomes allowed!
In Model I, since
\begin{equation}
\Gamma(H^+\to \tau^+\nu)/\Gamma(H^+\to c\bar s) \simeq 1/N_C = 1/3,
\end{equation}
one has BR$(H^+\to \tau^+\nu) \leq$ 30\%.
What is the $tbH^+$ coupling strength?
One sees from eq. (1) that $\lambda_{tbH^+} = V_{tb} \cot\beta \lambda_t$.
From $b\to s\gamma$ one infers that $\cot\beta < 2$,
while $0.25 < \lambda_t < 0.5$ for this mass range.
We thus find \cite{Hou1994} that $\lambda_{tbH^+} \sim 0.5$ -- $1 \sim g$,
hence,
$t\to bH^+ \gg t \to bW^*$ in the mass range of $m_{H^+} < m_t < M_W$.

\subsection{Model II: $t\to bH^+$ (thought to be ruled out)}

In this case one replaces $\cot\beta$ by $-\tan\beta$
in the coefficient of the $m_b$ term.
This makes all the difference compared to Model I.
The $H^+$ contribution now adds constructively
to the $W$ boson contribution 
(as well as the large QCD correction term that arises at
leading log (LL) order).
What is interesting \cite{HWil} is the appearance of a 
{\it $\tan\beta$-independent} term that contributes 
for any $\tan\beta$ value.
This amounts to a strong enhancement effect that is always there,
leading to a stringent constraint from the observation of
$b\to s\gamma$ and $B\to K^*\gamma$ by CLEO.
The upshot, as stated by CLEO \cite{CLEO}, is that
$m_{H^+} > 300$ GeV.

At LL order, one has significant scale dependence,
which is supposedly resolved at next-to-leading (NLL) order.
Such a calculation is rather tedious and intricate.
A partial calculation suggests that new cancellation effects
emerge at NLL order.
The details cannot be presented here \cite{HTGT},
but depending on
the sign of new NLL terms,
it may be possible to evade
$b\to s\gamma$ bound for $m_t < M_W$ and
$0.6 < \tan\beta < 1$.
At the same time, the right hand side of eq. (2)
gets multiplied by $m_\tau^2/(m_s^2 + \cot^4\beta m_c^2)$.
Thus, one needs $\tan\beta < 1$ to evade CDF direct
search for $t\to bH^+$.
The upshot, then,
is that $m_H^+ < m_t < M_W$ is in fact possible
if $\tan\beta \sim 1$, which is a value where the distinction
between Model I and II are blurred.

\subsection{General 2HDM: $t\to ch^0$}

Without NFC condition, in general one has FCNC
neutral Higgs boson couplings at tree level.
However, Nature seems to
have ``naturally" implemented some approximate flavor symmetry,
which is reflected in the mass and mixing hierarchies.
This suggests \cite{CS} that low energy FCNC constraints
could still be effectively evaded, {\it without}
constraining the high energy behavior.
The basic observation here then is that \cite{Hou},
there is in fact almost no {\it direct}
constraint that forbids FCNC $t$-$c$-scalar couplings.
It is NOT against any principle or experimental result \cite{HW}.
Assuming that some FCNC neutral scalar $h^0$ is rather light,
but still decaying via $b\bar b$,
then $t\to ch^0$ (followed by $h^0 \to b\bar b$) decay
provides a somewhat exotic but otherwise perfectly possible
scenario where $m_t < M_W$ could be realized \cite{Hou1994}.

\section{Singlet Charge 2/3 Quark and $R_b$ - $R_c$ Problem}

Another way to induce FCNC is to break GIM mechanism.
The easiest way beyond SM3 is the addition of
left-right singlet charge 2/3 quarks $Q_L$ and $Q_R$,
which affects top physics via $t$-$Q$ mixing.
One can show \cite{HH} that physical $m_u$ and $m_c$ eigenvalues
could remain small even with large $u$-$Q$ and $c$-$Q$
mixings.
Since $tcH$ coupling is induced by the
presence of {\it both} $c$-$Q$ and $t$-$Q$ mixings,
to allow $t\to cH$ decay
to dominate over $t\to W^*$ one needs both mixings to be very large.
However, on closer inspection, one would find that
too large a $c$-$Q$ mixing would lead to too small a
$Zcc$ coupling.
Interestingly, it has recently been reported
that the $Zcc$ coupling is 2.5$\sigma$ below SM3 expectations,
while $Zbb$ coupling seems to be almost 4$\sigma$ above.
Taking these experimental indications as hints,
it is possible to construct \cite{BBH} a solution to the so-called
$R_b$ -- $R_c$ problem (the only seeming discrepancy with
SM3 in town) in this context, at the price of
introducing a second Higgs doublet.
The latter provides for additional weak doublet splitting
($\Delta T$) via $m_{H^+} > v \gg m_{h^0}$, while the
exotic neutral Higgs $h^0$should be as light as possible to
allow maximal phase space for facilitating the hiding
of the actual top quark below $M_W$.
The dominantly singlet quark $Q$ emerges
as the heavy quark observed at the Tevatron.
It still dominantly decays via $bW$ mode,
but also decays via $sW$, $cZ$, $tZ$, $ch^0$, $th^0$,
with many interesting consequences \cite{BBH}.

The scenario bears some similarity with, but is also distinct from,
the MSSM solutions to the $R_b$ (but not $R_c$)
problem, where one demands light ``stop" and chargino/neutralinos.

\section{Fourth Generation Scenarios}

Some recent work illustrates that
a fourth generation could be entertained
from a high scale perspective.
Elaborate studies have been conducted
with SUSY GUT inspired structures.
For example, the work of Gunion, McKay and Pois \cite{GMP}
takes top (partner of $b$) to be the quark discovered
at Tevatron.
They then demonstrate that it is possible to have
$m_{b^\prime} < m_{t^\prime} < 120$ GeV,
where both heavy quarks remain unobserved if
$t^\prime \to b^\prime W^*$, where $W^*$ decay is rather soft,
while $b^\prime$ decays via 
loop induced $b^\prime \to bH$ mode \cite{bp}.
The discussion is rather elaborate, but the point,
to some extent, is to allow for a fourth generation but
also at the price of additional ``light" new particles,
which are again charged or neutral scalar bosons.
In a similar framework, 
the work of Carena, Haber and Wagner \cite{CHW}
suggests that $m_t \cong M_W$ via neutralino induced decays
$t\to \tilde t\chi^0$ (while $\tilde t \to c \chi^0$).
The fourth generation $t^\prime$ quark decays via standard
$bW$ channel and is the one observed at Tevatron.
One again has light new particles below $M_W$.

It may be worthwhile to strip ourselves from high scale
prejudices and reassess the issue of possible physics 
of a fourth generation at Tevatron and LEP II.

\section{Conclusion}

We have seen many scenarios where 
one may have particles below the $W$ scale.
In all cases one needs some light scalar particle
to facilitate their hiding.
We conclude that ``light" heavy quarks, below $M_W$,
or at least below $m_t$, is quite an open topic
for experimental study.

\vskip 0.3cm 
\noindent{\bf Acknowledgement}.
This work is supported in part by grant NSC 85-2112-M-002-011
of the Republic of China.
I wish to thank G.  Bhattacharyya, G. C. Branco,
C. Q. Geng, H. C. Huang, E. B. Tsai and P. Turcotte
for collaborative work.

\bibliographystyle{unsrt}

\vskip 0.9cm
\noindent {\Large\bf Figures}
\begin{description}
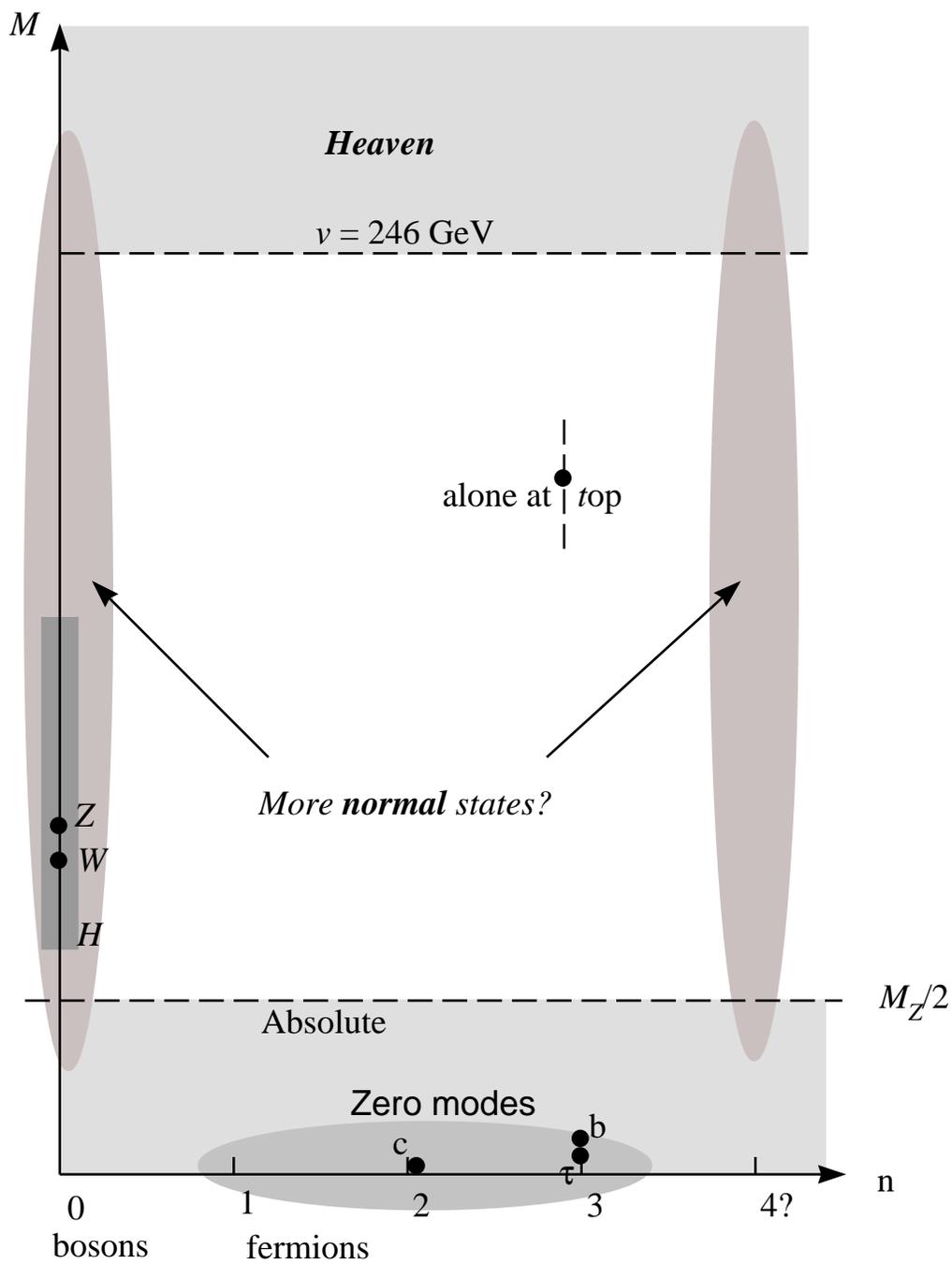
\item[Fig. 1.] Illustration of mass scales.
The ``usual" quarks and leptons appear as ``zero modes",
while $W$, $Z$, $H$ and the top quark appear
as ``normal" states on the $v$ scale,
suggesting the existence of many more such particles.

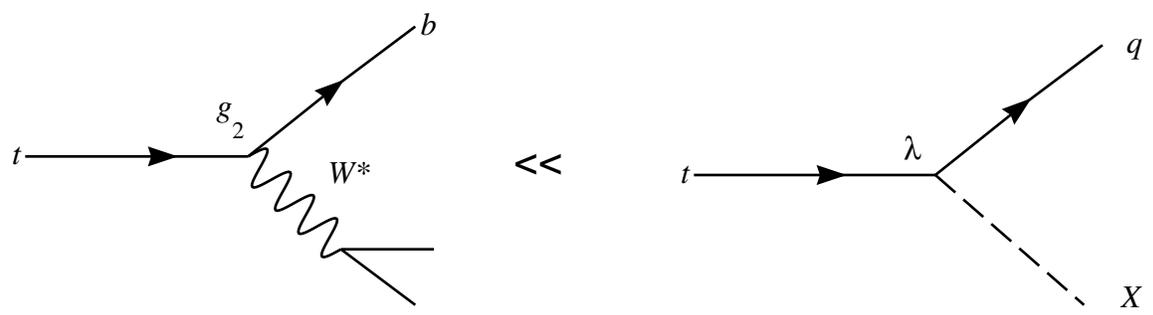
\item[Fig. 2.] Feynman diagrams to illustrate mechanism for
hiding top quark below $M_W$.

\end{description}

\end{document}